\documentclass[showpacs,preprintnumbers,prd,nofootinbib,floats,amssymb,floatfix]{revtex4-2}
\usepackage{amsmath}
\usepackage{graphicx}
\usepackage{amsxtra}
\usepackage{hyperref}
\usepackage{amssymb}
\usepackage{amstext}
\begin{document}

\title{Analysis of linearized Weyl gravity via the Hamilton-Jacobi method}
\author{Alberto Escalante}  \email{aescalan@ifuap.buap.mx}
\author{Víctor Alberto Zavala-P\'erez}  \email{vzavala@ifuap.buap.mx}
\affiliation{Instituto de F\'isica, Benem\'erita Universidad Aut\'onoma de Puebla. \\ Apartado Postal J-48 72570, Puebla Pue., M\'exico, }
 
\begin{abstract}
The Hamilton-Jacobi formalism is used to analyze the Weyl theory in the weak-field limit. The complete set of involutive Hamiltonians is obtained, which are classified into involutive and non-involutive. The counting of degrees of freedom is performed. Additionally, the generalized brackets and gauge symmetries are reported. 
\end{abstract}
\date{\today}
\preprint{}
\maketitle
\section{Introduction}{
Einstein's theory of general relativity (GR) is one of the most successful physical theories of all time \cite{1}; however, like any physical theory, it's not free of limitations. Perhaps the most important of them is that it is not renormalizable \cite{2}, which represents one of the great challenges of current physics \cite{3}. There are also some issues regarding the cosmological constant problem \cite{4} and the Ad hoc postulation of dark matter \cite{5} that compel us to look at alternatives and/or generalizations of GR. Within these efforts are the so-called higher-order theories. Higher-order theories are characterized by the presence of time derivatives of third  order or higher in the Lagrangian, which leads to equations of motion of at least fourth order, providing these theories with rich dynamics. Among these systems, those that contain squared  products of the curvature tensor are noteworthy, since these have been proven to be renormalizable \cite{6, 7} as well as improving the ultraviolet behavior of GR \cite{8}. Furthermore, these theories appear in fields such as string theory \cite{9, 10}, electrodynamics \cite{11, 12}, relativistic particles \cite{13, 14}, contact mechanics \cite{15}, as well as autonomous and non-autonomous dynamic systems \cite{16}. Unfortunately, higher-order theories do have some drawbacks, additional degrees of freedom and ghosts \cite{17} being the most notable among them. These ghosts are states with negative energy norms, which make the theory non-unitary. Despite the complications that this may present, interest in this type of system has not dwindled \cite{18, 19, 20}; with works such as those by Bender  \cite{21}, de la Cruz-Dombriz \cite{22}, and Paul \cite{23} showing that the ghost problem could be avoided. \\
The subject of this study is the higher-order theory known as Weyl gravity, which possesses, in addition to diffeomorphism covariance, invariance under conformal transformations of the metric. These angle-preserving scale transformations are an extension of Poincaré transformations, which include scale transformations \cite{24}. This theory has 6 degrees of freedom, associated with ordinary massless excitations of spin 2 and 1, as well as a spin 2 ghost \cite{25}. Weyl gravity has also been proven to be renormalizable \cite{19} and provides a natural explanation for dark energy \cite{18, 26} and it's solutions generalize the Schwarzschild solution, thus removing the need to invoke dark matter \cite{27,28}. That being said, it is still an open question whether these solutions are physically feasible \cite{29}.\\
Furthermore, the analysis of higher order theories cannot be performed by the standard procedures, so specialized techniques have been developed. The most common among these is the Ostrogradski-Dirac method \cite{23}. In this approach, both the fields and their corresponding time derivatives are taken as independent canonical variables, so that conjugate momenta are associated to them; thus extending the phase space. Then, the  constraints must be identified and classified, as the procedure has revealed the system as one with gauge symmetries. These are classified into first and second class, with the first-class being the generators of gauge transformations \cite{30}. This procedure usually encumbers the analysis, in which the constraints are sometimes chosen by hand in order to obtain a consistent algebra \cite{31}. \\
An alternative scheme is the Hamilton-Jacobi framework [HJ], which naturally handles the symmetries of the system in an elegant manner \cite{32}. In this approach, the order of the time derivatives is reduced by introducing additional degrees of freedom, which endows the system with constraints and highlights the existence of gauge symmetries. The constraints of the system are called Hamiltonians, classified as involutive or non-involutive, and treated on the same footing as the canonical Hamiltonian, yielding a system with several independent variables. The identification of the Hamiltonians is performed through the null vectors, an efficient  procedure that removes the need to fix the constraints by hand; which is otherwise commonplace in the usual methods \cite{12}. The integrability of the system is ensured by Frobenius integrability conditions, dealing with the non-involutive Hamiltonians by introducing the generalized brackets. At the end of the procedure, one can eliminate the additional degrees of freedom, retaining only the original ones. \\
With all said above, in this work the study of the Weyl action is performed via the Hamilton-Jacobi method, where the metric is perturbed around the Minkowski background. The order of the time derivatives of the Lagrangian is reduced by introducing an extrinsic curvature-like variable. The conjugate momenta are obtained via the usual definition, as well as the canonical Hamiltonian. The correct identification of the Hamiltonians is performed by means of the rank-nullity analysis of the Hessian matrix of the system, which are then classified by means of the Frobenius integrability conditions. The non-involutive Hamiltonians are integrated to the dynamics by introducing the generalized brackets, which generate a final set of involutive Hamiltonians from which the symmetries of the theory are obtained. This document uses the mostly positive sign convention of the metric $(-,+,+,+)$. Greek alphabet is used for space-time indices, whereas the Latin alphabet is used for spatial indices. A Hamiltonian analysis of Weyl's conformal gravity can also be found in \cite{33}, where a reduced Dirac formalism was used. We present an alternative to said analysis in which all fields will be  considered as dynamic. This includes both an extrinsic curvature-type variable and a Lagrange multiplier, the latter of which is usually taken as a momentum from the outset. Furthermore, a detailed HJ analysis of the gauge transformations is also presented, which in Dirac method  is usually shown only marginally.
}
\section{The Hamilton-Jacobi analysis}{
Weyl's action can be written as a squared function of the Riemann and Ricci tensors and the scalar curvature \cite{34}
\begin{equation}
S = \int \sqrt{-g}(R_{\alpha\beta\mu\nu}R^{\alpha\beta\mu\nu} - 2R_{\mu\nu}R^{\mu\nu} + \frac{1}{3}R^{2})d^{4}x.
\label{eq01}
\end{equation}
We begin the analysis by making a perturbative expansion of the metric
\begin{equation}
g_{\mu\nu} = \eta_{\mu\nu} + h_{\mu\nu},
\label{eq02}
\end{equation}
which yields the linearized Weyl action, shown here for completeness
\begin{align}
S &=\int\Big[- \frac{1}{2}\partial_{\alpha}\partial_{\beta}h_{\gamma}{}^{\alpha}\partial_{\delta}\partial^{\delta}h^{\beta\gamma} + \frac{1}{2}\partial_{\alpha}\partial_{\beta}h_{\gamma}{}^{\alpha}\partial^{\gamma}\partial_{\delta}h^{\beta\delta} + \frac{1}{4}\partial_{\alpha}\partial^{\alpha}h_{\beta\gamma}\partial_{\delta}\partial^{\delta}h^{\beta\gamma} \nonumber \\[5pt]
&- \frac{1}{12}\partial_{\alpha}\partial_{\beta}h\partial^{\alpha}\partial^{\beta}h - \frac{1}{3}\partial^{\alpha}\partial^{\beta}h_{\alpha\beta}\partial^{\gamma}\partial^{\delta}h_{\gamma\delta} + \frac{1}{6}\partial^{\alpha}\partial^{\beta}h_{\alpha\beta}\partial_{\gamma}\partial^{\gamma}h\Big]d^{4}x.
\label{eq03}
\end{align}
We perform a $3+1$ decomposition and reduce the order of the time-derivatives by introducing the following change of variable
\begin{equation}
K_{ij} = \frac{1}{2}(\dot{h}_{ij} - \partial_{i}h_{0j} - \partial_{j}h_{0i}),
\label{eq04}
\end{equation}
obtaining the following form of the action
\begin{align}
\mathcal{L} &= \frac{1}{2}(\dot{K}_{ij}\dot{K}^{ij} + \tilde{R}_{ij}\tilde{R}^{ij} + \dot{K}_{ij}\partial^{i}\partial^{j}h^{00} - 2\dot{K}_{ij}\tilde{R}^{ij} - \partial_{i}\partial_{j}h_{00}\tilde{R}^{ij}) - \frac{1}{6}(\dot{K}^{2} \nonumber \\[5pt] 
&+ \tilde{R}^{2} + \dot{K}\nabla^{2}h_{00} - 2\dot{K}\tilde{R} - \tilde{R}\nabla^{2}h_{00}) + \frac{1}{12}\nabla^{2}h_{00}\nabla^{2}h_{00} - 2\partial_{j}K_{ik}\partial^{j}K^{ik} \nonumber \\[5pt]
&+ 3\partial^{i}K_{ij}\partial_{k}K^{kj} - 2\partial_{i}K\partial_{j}K^{ij} + \partial_{i}K\partial^{i}K - \lambda^{ij}\Big[K_{ij} - \frac{1}{2}(\dot{h}_{ij} - h_{0j,i} - h_{0i,j})\Big],
\label{eq05}
\end{align}
where  contraction with space indices is expressed by a tilde $\tilde{R} = R_{i}{}^{i}$ . Notice how the theory is now of first order, and that the last term is enforcing the change of variable via a Lagrange multiplier $\lambda^{ij}$, which will be  taken as an additional dynamical variable. It is worth commenting that in HJ framework the canonical variables are all treated on the same footing, thus, all of them are assigned their canonical moments. This approach is different from that reported in \cite{33}, where the canonical momenta are associated only to  the fields with time derivative. Therefore, the system's variables are $\mathcal{Q}^{(\mu)} = \{K_{ab}, h_{ab}, h_{0a}, h_{00}, \lambda^{ab}\}$. We can find their conjugate momenta $\mathcal{P}^{(\mu)} = \{\Pi^{ab}, \pi^{ab}, \pi^{0a}, \pi^{00}, \Upsilon_{ab}\}$ via the usual definition
\begin{equation}
\mathcal{P}^{(\mu)} = \frac{\partial \mathcal{L}}{\partial \mathcal{\dot{Q}}^{(\mu)}}.
\label{eq06}
\end{equation}
These are
\begin{align}
\Pi^{ab} &= \dot{K}^{ab} + \frac{1}{2}h^{00,a,b} - \tilde{R}^{ab} - \frac{1}{3}\eta^{ab}\dot{K} - \frac{1}{6}\eta^{ab}\nabla^{2}h_{00} + \frac{1}{3}\eta^{ab}\tilde{R}, \label{eq07} \\[5pt]
\pi^{ab} &= \frac{1}{2}\lambda^{ba}, \label{eq08} \\[5pt]
\pi^{0a} &= \frac{\partial \mathcal{L}}{\partial \dot{h}_{0a}} = 0, \label{eq09} \\[5pt]
\pi^{00} &= \frac{\partial \mathcal{L}}{\partial \dot{h}_{00}} = 0, \label{eq10} \\[5pt]
\Upsilon_{ab} &= \frac{\partial \mathcal{L}}{\partial \dot{\lambda}^{ab}} = 0. \label{eq11} 
\end{align}
Taking into account their symmetries they satisfy the expected canonical relations
\begin{align}
\{K_{ij}(x), \Pi^{ab}(y)\} &= \frac{1}{2}\left(\delta_{i}^{a}\delta_{j}^{b} + \delta_{j}^{a}\delta_{i}^{b}\right)\delta^{3}(x-y), \label{eq12} \\[5pt]
\{h_{ij}(x), \pi^{ab}(y)\} &= \frac{1}{2}\left(\delta_{i}^{a}\delta_{j}^{b} + \delta_{j}^{a}\delta_{i}^{b}\right)\delta^{3}(x-y), \label{eq13} \\[5pt]
\{h_{0i}(x), \pi^{0a}(y)\} &= \delta_{i}^{a}\delta^{3}(x-y), \label{eq14} \\[5pt]
\{h_{00}(x), \pi^{00}(y)\} &= \delta^{3}(x-y), \label{eq15} \\[5pt]
\{\lambda^{ij}(x), \Upsilon_{ab}(y)\} &= \frac{1}{2}\left(\delta_{a}^{i}\delta_{b}^{j} + \delta_{b}^{i}\delta_{a}^{j}\right)\delta^{3}(x-y). \label{eq16} 
\end{align}
The canonical Hamiltonian is obtained with a Legendre transformation using the complete set of variables $\mathcal{Q}^{(\mu)}$
\begin{align}
\mathcal{H}_{c} &= \dot{h}_{ij}\pi^{ij} + \dot{h}_{0i}\pi^{0i} + \dot{h}_{00}\pi^{00} + \dot{K}_{ij}\Pi^{ij} + \dot{\lambda}_{ij}\Upsilon^{ij} - \mathcal{L}.
\label{eq17}
\end{align}
Which takes the following form
\begin{align}
\mathcal{H}_c &= \frac{1}{2}\Pi_{ij}\Pi^{ij} - \frac{1}{2}\Pi_{ij}\partial_{i}\partial_{j}h^{00} + \Pi_{ij}\tilde{R}^{ij} + 2\partial_{j}K_{ik}\partial^{j}K^{ik} \nonumber \\[5pt]
&- 3\partial^{i}K_{ij}\partial_{k}K^{kj} + 2\partial_{i}K\partial_{j}K^{ij} - \partial_{i}K\partial^{i}K + 2\pi^{ij}\left(K_{ij} + h_{0i,j}\right), 
\label{eq18}
\end{align}
where we can observe linear terms in the momenta, associated to  Ostrogradki's instabilities. Reducing the order in a system as we have done introduces Hamiltonians between the variables, which have to be accounted for. These Hamitonians can be involutive or non-involutive; the former are those whose Poisson brackets with all Hamiltonians (including themselves) are zero, otherwise they are called non-involutive. The correct number of Hamiltonians  is obtained via the nullity of the system's Hessian matrix, which can be easily seen to be 17. Equations (\ref{eq07}-\ref{eq11}) account for all the constraints but one. Which is obtained trivially from the Hessian matrix as the contraction of (\ref{eq07}) with the null vector, say, $v^\mu=(\eta_{ij}, 0, 0, 0, 0)$. The Hamiltonians  are
\begin{align}
&\mathcal{H}: \mathcal{H}_c +{\pi}=0, &&\Omega_{(0)}:\Pi = 0, &&\Omega_{(1)} : \pi^{00} = 0, &&\Omega_{(2)}^{0i} : \pi^{0i} = 0, \nonumber \\[5pt]
&\Omega_{(3)}^{ij} :-2\pi^{ij} +\lambda^{ij} = 0, &&\Omega^{(4)}_{ij} : \Upsilon_{ij} = 0,  &&
\label{eq19}
\end{align}
where $\mathcal{H}_c$ is the canonical hamiltonian and ${\pi}=\partial_0S$. In the HJ framework, the dynamics are determined by the fundamental differential, defined as
\begin{equation}
dF = \{F,\mathcal{H}_{(\mu)}\}d\xi^{(\mu)},
\end{equation}
where $F$ is a function of the phase space variables, $\xi^{(\mu)}$ are evolution parameters related to $\mathcal{H}_{\mu}$, which are the Hamiltonians of the system. These Hamiltonians must form a complete set, and indeed a null vector analysis will trivially show that they are. They also need to be independent from each other to ensure a consistent dynamics, so we impose Frobenius integrability conditions upon them. This amounts to take the Poisson bracket between the Hamiltonians as well as having a closed algebra. The only non-trivial bracket is as follows
\begin{equation}
\{\Omega_{(3)}^{ij}(x), \Omega^{(4)}_{ab}(y)\} = \frac{1}{2}\left(\delta_{a}^{i}\delta_{b}^{j} + \delta_{b}^{i}\delta_{a}^{j}\right)\delta^{3}(x-y),
\label{eq20}
\end{equation}
which means that both $\Omega_{(3)}$ and $\Omega^{(4)}$ are non-involutive, from now on they will be labeled as $\Lambda_{(3)}$ and $\Lambda^{(4)}$ respectively. These are removed via the introduction of the generalized bracket
\begin{equation}
\{A(x), B(x^{\prime})\}^{*} = \{A(x),B(x^{\prime})\} - \iint \{A(x),\Lambda^{(\mu)}(y)\}\mathbf{\Delta}_{(\mu)}^{(\nu)^{-1}}(y,z)\{\Lambda_{(\nu)}(z), B(x^{\prime})\} d^{2}y d^{2}z,
\label{eq21}
\end{equation}
where $\mathbf{\Delta}_{(\mu)}^{(\nu)}$ is a matrix, whose entries are Poisson brackets between the non-involutive constraints
\begin{equation}
\mathbf{\Delta}_{(\nu)}^{(\mu)} = 
\begin{pmatrix}
\{\Lambda^{1}, \Lambda^{1}\} & \{\Lambda^{1}, \Lambda^{2}\} & \cdots & \{\Lambda^{1}, \Lambda^{n}\} \\
\{\Lambda^{2}, \Lambda^{1}\} & \{\Lambda^{2}, \Lambda^{2}\} & \cdots & \{\Lambda^{2}, \Lambda^{n}\} \\
\vdots  & \vdots  & \ddots & \vdots  \\
\{\Lambda^{n}, \Lambda^{1}\} & \{\Lambda^{n}, \Lambda^{2}\} & \cdots & \{\Lambda^{n}, \Lambda^{n}\} 
\end{pmatrix}.
\label{eq22}
\end{equation}
Since we only have two non-involutive Hamiltonians  we expect a $2 \times 2$ antisymmetric matrix, given by  
\begin{equation}
\mathbf{\Delta}^{ij}_{ab}(x,y) = \frac{1}{2}(\delta_{a}^{i}\delta_{b}^{j} + \delta_{b}^{i}\delta_{a}^{j})\begin{pmatrix} 0 & 1  \\-1 & 0  \end{pmatrix}\delta^{3}(x-y).
\label{eq23}
\end{equation}
Whose inverse can easily be found to be
\begin{equation}
\mathbf{\Delta}^{bk^{-1}}_{cj}(y,z) = \delta_{c}^{b}\delta_{j}^{k} \begin{pmatrix} 0 & -1 \\ 1 & 0 \end{pmatrix}\delta^{2}(y-z).
\label{eq24}
\end{equation}
We can obtain the new dynamics of the system by introducing this matrix back into (\ref{eq21}) and calculating the new brackets for our variables $\mathcal{Q}^{(\mu)}$ and $\mathcal{P}^{(\mu)}$
\begin{align}
&\{K_{ij}(x), \Pi^{kl}(x^{\prime})\}^{\ast} = \frac{1}{2}(\delta_{i}^{k}\delta_{j}^{l} + \delta_{i}^{l}\delta_{j}^{k})\delta^{3}(x-x^{\prime}), \label{eq25} \\[5pt]
&\{h_{ij}(x), \pi^{kl}(x^{\prime})\}^{\ast} = \frac{1}{2}(\delta_{i}^{k}\delta_{j}^{l} + \delta_{i}^{l}\delta_{j}^{k})\delta^{3}(x-x^{\prime}), \label{eq26} \\[5pt]
&\{h_{0i}(x), \pi^{0k}(x^{\prime})\}^{\ast} = \delta_{i}^{k}\delta^{3}(x-x^{\prime}), \label{eq27} \\[5pt]
&\{h_{00}(x), \pi^{00}(x^{\prime})\}^{\ast} = \delta^{3}(x-x^{\prime}), \label{eq28} \\[5pt]
&\{\Upsilon_{ij}(x), \lambda^{kl}(x^{\prime})\}^{\ast} = 0, \label{eq29}  \\[5pt]
&\{h_{ij}(x), \lambda^{kl}(x^{\prime})\}^{\ast} = (\delta_{i}^{k}\delta_{j}^{l} + \delta_{i}^{l}\delta_{j}^{k})\delta^{3}(x-x^{\prime}). \label{eq30} 
\end{align}
These equations will replace (\ref{eq12})-(\ref{eq16}) as the new canonical relationships. Note that only relation (\ref{eq16}) has been modified, which is now (\ref{eq29}); in addition, the relation (\ref{eq30}) has been added to the set. These are the only two modifications to the original Poisson brackets. Before proceeding further, we would like to make two observations on the differences between these and the initial algebra. First, $\Upsilon_{ij}$, which was introduced as the conjugate momentum to $\lambda^{ij}$, is now fully decoupled from the system. Second is that $\lambda^{ij}$ can be treated like a conjugate momentum to $h_{ij}$ by virtue of (\ref{eq30}). Both of these properties happen as a consequence of the generalized bracket, which has now reduced   the system, allowing us to remove the non-involutive Hamiltonians from the fundamental differential. 
Removing  the non-involutive Hamiltonians  leaves us with only the involutive ones 
\begin{equation}
\Omega_{(0)} = \tilde{\Pi}, \quad  \quad \Omega_{(1)} = \pi^{00}, \quad  \quad \Omega_{(2)}^{0i} = \pi^{0i}, 
\label{eq32}
\end{equation} 
this means that the fundamental differential is given by
\begin{equation}
dF = \int\Big[\{F, \mathcal{H}\}^{\ast}dt + \{F, \Omega_{(0)}\}^{\ast}d\xi^{(0)} + \{F, \Omega_{(1)}\}^{\ast}d\xi^{(1)} + \{F, \Omega_{(2)}^{0i}\}^{\ast}d\xi_{0i}^{(2)}\Big]d^{3}y.
\label{eq32a}
\end{equation}
Since the generalized bracket has modified the structure of the phase space we must revisit the Frobenius integrability conditions; that is, we must insist that the Hamiltonians in (\ref{eq32}) are still involutive under the generalized bracket (\ref{eq32a}). This leads to 
\begin{align}
d\Omega_{(0)} &= \int\left[\{\Omega_{(0)}(x), \mathcal{H}(y)\}^{*}dt + \{\Omega_{(0)}(x), \Omega_{(\mu)}(y)\}^{*}d\xi^{(\mu)}\right]d^{3}y = - 2\tilde{\pi}dt \label{eq33} \\[5pt]
d\Omega_{(1)} &= \int\left[\{\Omega_{(1)}(x), \mathcal{H}(y)\}^{*}dt + \{\Omega_{(1)}(x), \Omega_{(\mu)}(y)\}^{*}d\xi^{(\mu)}\right]d^{3}y  = \frac{1}{2}\partial_{i}\partial_{j}\Pi^{ij}dt \label{eq34} \\[5pt]
d\Omega_{(2)}^{0i} &= \int\left[\{\Omega_{(2)}^{0i}(x), \mathcal{H}(y)\}^{*}dt + \{\Omega_{(2)}^{0i}(x), \Omega_{(\mu)}(y)\}^{*}d\xi^{(\mu)}\right]d^{3}y = 2\partial_{j}\pi^{ij}dt \label{eq35}
\end{align}
Here, the index $(\mu)$ accounts all the Hamiltonians in (\ref{eq32}). For the system to be in involution these expressions must be either zero or a combination of Hamiltonians. For this reason, these expressions must be added as new Hamiltonians, which naturally have to fulfill the Frobenius integrability conditions as well. The new set of Hamiltonians $\Omega^{(\mu)}$ is 
\begin{align} 
&\Omega_{(0)} = \tilde{\Pi}, &&\Omega_{(1)} = \pi^{00}, &&\Omega_{(2)}^{0i} = \pi^{0i}, \nonumber \\[5pt]
&\Omega_{(6)} = \tilde{\pi}, &&\Omega_{(7)} = \partial_{i}\partial_{j}\Pi^{ij}, &&\Omega_{(8)}^{0i} = \partial_{j}\pi^{ij}.
\label{eq36}
\end{align}
Since all these expressions are functions of the canonical moments, they are trivially in involution with themselves. Again, only the generalized brackets between $\Omega_{(6)}$, $\Omega_{(7)}$, and $\Omega_{(8)}^{0i}$ with the  Hamiltonian $\mathcal{H}$ are to be be examined
\begin{align}
\int\{\Omega_{(6)}(x),\mathcal{H}(y)\}^{\ast}d^{3}y &= \frac{1}{2}\nabla^{2}\Omega_{(0)} + \frac{1}{2}\Omega_{(6)} = 0, \label{eq37} \\[5pt]
\int\{\Omega_{(7)}(x), \mathcal{H}(y)\}^{\ast}d^{3}y &= 2\partial_{i}\Omega^{0i}_{(7)} = 0, \label{eq38} \\[5pt]
\int\{\Omega_{(7)}^{0i}(x), \mathcal{H}(y)\}^{\ast}d^{3}y &= 0. \label{eq39}
\end{align}
All these equations vanish, so the system is in complete involution. The final set of Hamiltonians $\Omega^{(\mu)}$ is that of (\ref{eq36}), which together with the  Hamiltonian $\mathcal{H}$ will generate the dynamics of the system. This is done via the fundamental differential
\begin{align}
dF &= \int\Big[\{F, \mathcal{H}\}^{\ast}dt + \{F, \Omega_{(0)}\}^{\ast}d\xi^{(0)} + \{F, \Omega_{(1)}\}^{\ast}d\xi^{(1)} + \{F, \Omega_{(2)}^{0i}\}^{\ast}d\xi_{0i}^{(2)} \nonumber \\[5pt]
&+ \{F, \Omega_{(6)}\}^{\ast}d\xi^{(6)} + \{F, \Omega_{(7)}\}^{\ast}d\xi^{(7)} + \{F, \Omega^{0i}_{(8)}\}^{\ast}d\xi_{0i}^{(8)}\Big]d^{3}y
\label{eq40}
\end{align}
Where $\xi^{(\mu)}$ are evolution parameters associated with each of the Hamiltonians $\Omega_{(\mu)}$. The evolution with respect to one of these parameters, including time, is independent of the others by virtue of the Frobenius integrability conditions. Thus, equation (\ref{eq40}) has all the information of the dynamics of the system, including gauge transformations, which will be shown below. By applying the fundamental differential to the variables we first get the HJ characteristic equations
\begin{align}
dh_{ij} &= (2K_{ij} + \partial_{i}h_{0j}+ \partial_{j}h_{0i})dt + \eta_{ij}d\xi^{(6)} - \frac{1}{2}(\partial_{j}d\xi_{0i}^{(8)} + \partial_{i}d\xi_{0j}^{(8)}), \label{eq41} \\[5pt]
dh_{0i} &= d\xi_{0i}^{(2)}, \label{eq42} \\[5pt]
dh_{00} &= d\xi^{(1)}, \label{eq43} \\[5pt]
d\lambda^{ij} &= -\big[\partial^{i}\partial_{m}\Pi^{mj} - \nabla^{2}\Pi^{ij} + \partial^{j}\partial_{m}\Pi^{mi} \big]dt, \label{eq44} \\[5pt]
dK_{ij} &= (\Pi_{ij} - \frac{1}{2}\partial_{i}\partial_{j}h_{00} + \tilde{R}_{ij})dt + \eta_{ij}d\xi^{(0)} + \partial_{i}\partial_{j}d\xi^{(7)} \label{eq45} \\[5pt]
d\Upsilon_{ij} &= 0, \label{eq46} \\[5pt]
d\pi^{00} &= 0, \label{eq47} \\[5pt]
d\pi^{0i} &= 0, \label{eq48} \\[5pt]
d\pi^{ij} &= - \frac{1}{2}\Big[\partial^{i}\partial_{m}\Pi^{mj} - \nabla^{2}\Pi^{ij} + \partial^{j}\partial_{m}\Pi^{mi}\Big]dt, \label{eq49} \\[5pt]
d\Pi^{ij} &= \Big[4\nabla^{2}K^{ij} - 3\partial^{i}\partial_{m}K^{mj} - 3\partial^{j}\partial_{m}K^{mi} + 2\partial^{i}\partial^{j}K + 2\eta^{ij}\partial^{m}\partial^{n}K_{mn} \label{eq50} \\[5pt]
&- 2\eta^{ij}\nabla^{2}K - 2\pi^{ij}\Big]dt. \label{eq51}
\end{align}
We can see that not only are $\Upsilon_{ij}$, $\pi^{00}$, and $\pi^{0i}$ not dynamical, but also that (\ref{eq44}) and (\ref{eq49}) give the same information. Taking the parameters $\xi^{(\mu)} = 0$ we are left with only the following equations
\begin{align}
dh_{ij} &= (2K_{ij} + \partial_{i}h_{0j}+ \partial_{j}h_{0i})dt, \label{eq52} \\[5pt]
dK_{ij} &= (\Pi_{ij} - \frac{1}{2}\partial_{i}\partial_{j}h_{00} + \tilde{R}_{ij})dt, \label{eq53} \\[5pt]
d\pi^{ij} &= - \frac{1}{2}\Big[\partial^{i}\partial_{m}\Pi^{mj} - \nabla^{2}\Pi^{ij} + \partial^{j}\partial_{m}\Pi^{mi}\Big]dt, \label{eq54} \\[5pt]
d\Pi^{ij} &= \Big[4\nabla^{2}K^{ij} - 3\partial^{i}\partial_{m}K^{mj} - 3\partial^{j}\partial_{m}K^{mi} + 2\partial^{i}\partial^{j}K + 2\eta^{ij}\partial^{m}\partial^{n}K_{mn} \label{eq55} \\[5pt]
&- 2\eta^{ij}\nabla^{2}K - 2\pi^{ij}\Big]dt. \label{eq56}
\end{align}
The first of these is equation (\ref{eq04}), which when combined with the second gives us the equation of motion for $h_{ij}$, so the number of dynamical variables is $18$. Not taking into account $\Omega^{0i}_{(1)}$ and $\Omega^{0i}_{(2)}$ (since $\pi^{00}$ and $\pi^{0i}$ are not dynamical) we have six first class constraints, therefore the system has six degrees of freedom. \\
The canonical transformations can be determined by taking $dt = 0$ in the characteristic equations and working exclusively with the relations  that the additional parameters $\xi^{(\mu)}$ provide
\begin{align}
dh_{ij} &= \eta_{ij}d\xi^{(6)} - \frac{1}{2}(\partial_{j}d\xi_{0i}^{(8)} + \partial_{i}d\xi_{0j}^{(8)}), \label{eq57} \\[5pt]
dh_{0i} &= d\xi_{0i}^{(2)}, \label{eq58} \\[5pt]
dh_{00} &= d\xi^{(1)}, \label{eq59} \\[5pt]
dK_{ij} &= \eta_{ij}d\xi^{(0)} + \partial_{i}\partial_{j}d\xi^{(7)}. \label{eq60}
\end{align}
The gauge symmetries of the theory can now be obtained by making the action invariant under the variations of the metric perturbation (\ref{eq57})-(\ref{eq59}) 
\begin{align}
\delta S &= \int\Big[- \square\partial^{\mu}\partial_{\sigma}h^{\sigma\nu} + \frac{1}{3}\partial^{\mu}\partial^{\nu}\partial_{\sigma}\partial_{\delta}h^{\sigma\delta} + \frac{1}{2}\square\square h^{\mu\nu} \nonumber \\[5pt]
&- \frac{1}{6}\eta^{\mu\nu}\square\square h_{\delta}{}^{\delta} + \frac{1}{6}\square\partial^{\mu}\partial^{\nu}h_{\delta}{}^{\delta} + \frac{1}{6}\eta^{\mu\nu}\square\partial^{\alpha}\partial^{\beta}h_{\alpha\beta}\Big]\delta h_{\mu\nu}dtd^{3}x
\label{eq61}
\end{align}
Which have to be grouped into a single covariant expression
\begin{equation}
\delta h_{\mu\nu} = \delta_{\mu}^{0}\delta_{\nu}^{0}\delta\xi^{(1)} + \frac{1}{2}(\delta_{\mu}^{0}\delta_{\nu}^{i} + \delta_{\mu}^{i}\delta_{\nu}^{0})\delta\xi^{(2)}_{0i} + \delta_{\mu}^{i}\delta_{\nu}^{j}(\eta_{ij}\delta\xi^{(6)} - \frac{1}{2}\partial_{j}\delta\xi^{(8)}_{0i} - \frac{1}{2}\partial_{i}\delta\xi^{(8)}_{0j}).
\label{eq62}
\end{equation}
Written in terms of the parameters $\xi^{(\mu)}$ the variation of the action becomes
\begin{align}
\delta S &= \int\Big[\Big(- \square\partial^{0}\partial_{\sigma}h^{\sigma 0} + \frac{1}{3}\partial^{0}\partial^{0}\partial_{\sigma}\partial_{\delta}h^{\sigma\delta} + \frac{1}{2}\square\square h^{00} - \frac{1}{6}\eta^{00}\square\square h + \frac{1}{6}\square\partial^{0}\partial^{0}h \nonumber \\[5pt]
&+ \frac{1}{6}\eta^{00}\square\partial^{\alpha}\partial^{\beta}h_{\alpha\beta}\Big)\delta\xi^{(1)} + \Big(- \frac{1}{2}\square\partial^{0}\partial_{\sigma}h^{\sigma i} + \frac{1}{3}\partial^{0}\partial^{i}\partial_{\sigma}\partial_{\delta}h^{\sigma\delta} + \frac{1}{2}\square\square h^{0i} \nonumber \\[5pt]
&+ \frac{1}{6}\square\partial^{0}\partial^{i}h - \frac{1}{2}\square\partial^{i}\partial_{\sigma}h^{\sigma 0}\Big)\delta\xi^{(2)}_{0i} + \Big(-\square\partial_{i}\partial_{\alpha}h^{\alpha i} + \frac{1}{3}\nabla^{2}\partial_{\sigma}\partial_{\delta}h^{\sigma\delta} + \frac{1}{2}\square\square \tilde{h} \nonumber \\[5pt]
&- \frac{1}{2}\square\square h + \frac{1}{6}\square\nabla^{2}h + \frac{1}{2}\square\partial^{\alpha}\partial^{\beta}h_{\alpha\beta}\Big)\delta\xi^{(6)} + \Big(\frac{1}{2}\square\partial^{i}\partial_{\sigma}h^{\sigma j} - \frac{1}{6}\partial^{i}\partial^{j}\partial_{\sigma}\partial_{\delta}h^{\sigma\delta} \nonumber \\[5pt]
&- \frac{1}{4}\square\square h^{ij} + \frac{1}{12}\eta^{ij}\square\square h - \frac{1}{12}\square\partial^{i}\partial^{j}h - \frac{1}{12}\eta^{ij}\square\partial^{\alpha}\partial^{\beta}h_{\alpha\beta}\Big)\left(\partial_{j}\delta\xi^{(8)}_{0i} + \partial_{i}\delta\xi^{(8)}_{0j}\right)\Big]dtd^{3}x.
\label{eq63}
\end{align}
After some algebraic manipulations, involving the expansion and contraction of indices, this expression can be rewritten as follows
\begin{equation}
\delta S = \int\Big[\Xi_{(1)}\delta\xi^{(1)} + \Xi^{i}_{(2)}\delta\xi_{0i}^{(2)} + \Xi_{(6)}\delta\xi^{(6)} + \Xi^{ij}_{(8)}\left(\partial_{j}\delta\xi^{(8)}_{0i} + \partial_{i}\delta\xi^{(8)}_{0j}\right)\Big]dtd^{3}x
\label{eq64}
\end{equation}
With
\begin{align}
\Xi_{(1)} &= \frac{2}{3}\nabla^{2}\partial_{0}\partial_{i}h^{0i} - \frac{1}{2}\partial_{0}\partial^{0}\partial_{i}\partial_{j}h^{ij} + \frac{1}{3}\nabla^{2}\nabla^{2}h^{00} + \frac{1}{6}\nabla^{2}\partial_{0}\partial^{0}\tilde{h} + \frac{1}{6}\nabla^{2}\nabla^{2}\tilde{h} - \frac{1}{6}\nabla^{2}\partial_{i}\partial_{j}h^{ij}, \label{eq65} \\[5pt]
\Xi^{i}_{(2)} &= \frac{1}{2}\partial_{0}\partial^{0}\partial_{0}\partial_{j}h^{ji} + \frac{1}{2}\nabla^{2}\partial_{0}\partial_{j}h^{ji} + \frac{1}{6}\partial^{0}\partial^{i}\partial_{0}\partial_{j}h^{0j} + \frac{1}{3}\partial^{0}\partial^{i}\partial_{j}\partial_{k}h^{jk} + \frac{1}{2}\nabla^{2}\partial_{0}\partial^{0}h^{0i} \nonumber \\[5pt]
&+ \frac{1}{2}\nabla^{2}\nabla^{2}h^{0i} + \frac{1}{6}\partial^{0}\partial^{i}\partial_{0}\partial^{0}\tilde{h} + \frac{1}{3}\nabla^{2}\partial^{0}\partial^{i}h^{00} + \frac{1}{6}\nabla^{2}\partial^{0}\partial^{i}\tilde{h} - \frac{1}{2}\nabla^{2}\partial^{i}\partial_{j}h^{j0}, \label{eq66} \\[5pt]
\Xi_{(6)} &= - \frac{1}{2}\partial_{0}\partial^{0}\partial_{i}\partial_{j}h^{ij} - \frac{1}{6}\nabla^{2}\partial_{i}\partial_{j}h^{ij} + \frac{2}{3}\nabla^{2}\partial_{0}\partial_{i}h^{0i} + \frac{1}{3}\nabla^{2}\nabla^{2}h^{00} + \frac{1}{6}\nabla^{2}\partial_{0}\partial^{0}\tilde{h} + \frac{1}{6}\nabla^{2}\nabla^{2}\tilde{h}, \label{eq67} \\[5pt]
\Xi^{ij}_{(8)} &= - \frac{1}{2}\partial^{0}\partial^{i}\partial_{0}\partial^{0}h^{0j} + \frac{1}{2}\partial_{0}\partial^{0}\partial^{i}\partial_{k}h^{kj} - \frac{1}{2}\nabla^{2}\partial^{0}\partial^{i}h^{0j} + \frac{1}{2}\nabla^{2}\partial^{i}\partial_{k}h^{kj} + \frac{1}{4}\partial_{0}\partial^{0}\partial^{i}\partial^{j}h^{00} \nonumber \\[5pt]
&- \frac{1}{3}\partial^{i}\partial^{j}\partial_{0}\partial_{k}h^{0k} - \frac{1}{6}\partial^{i}\partial^{j}\partial_{k}\partial_{l}h^{kl} - \frac{1}{4}\partial_{0}\partial^{0}\partial_{0}\partial^{0}h^{ij} - \frac{1}{2}\nabla^{2}\partial_{0}\partial^{0}h^{ij} - \frac{1}{4}\nabla^{2}\nabla^{2}h^{ij} \nonumber \\[5pt]
&+ \frac{1}{12}\eta^{ij}\partial_{0}\partial^{0}\partial_{0}\partial^{0}\tilde{h} + \frac{1}{12}\eta^{ij}\nabla^{2}\partial_{0}\partial_{0}h^{00} + \frac{1}{6}\eta^{ij}\nabla^{2}\partial_{0}\partial^{0}\tilde{h} - \frac{1}{12}\eta^{ij}\nabla^{2}\nabla^{2}h^{00} + \frac{1}{12}\eta^{ij}\nabla^{2}\nabla^{2}\tilde{h} \nonumber \\[5pt]
&- \frac{1}{12}\partial_{0}\partial^{0}\partial^{i}\partial^{j}\tilde{h} + \frac{1}{12}\nabla^{2}\partial^{i}\partial^{j}h^{00} - \frac{1}{12}\nabla^{2}\partial^{i}\partial^{j}\tilde{h} - \frac{1}{6}\eta^{ij}\partial_{0}\partial^{0}\partial_{0}\partial_{k}h^{0k} - \frac{1}{12}\eta^{ij}\partial_{0}\partial^{0}\partial_{k}\partial_{l}h^{kl} \nonumber \\[5pt]
&- \frac{1}{6}\eta^{ij}\nabla^{2}\partial_{0}\partial_{k}h^{0k} - \frac{1}{12}\eta^{ij}\nabla^{2}\partial_{k}\partial_{l}h^{kl}. \label{eq68}
\end{align}
After a brief inspection it can be seen that equations (\ref{eq65}) and (\ref{eq67}) are identical,  we rename them as follows 
$\Xi_{(1)} = \Xi_{(6)} \equiv \Xi$. This will relate the associated parameters $\delta\xi^{(1)}$ and $\delta\xi^{(6)}$. By making this substitution in (\ref{eq64}) and after some partial integrations we arrive to 
\begin{equation}
\delta S = \int\left[\Xi\left(\delta\xi^{(1)} + \delta\xi^{(6)}\right) + \Xi_{(2)}^{0i}\delta\xi_{0i}^{(2)} - \left(\partial_{j}\Xi_{(8)}^{ij} + \partial_{j}\Xi_{(8)}^{ji}\right)\delta\xi_{0i}^{(8)}\right]dtd^3x
\label{eq69}
\end{equation}
With
\begin{align}
\partial_{j}\Xi^{ij}_{(8)} + \partial_{j}\Xi^{ji}_{(8)} &= - \frac{1}{2}\nabla^{2}\partial^{0}\partial_{0}\partial^{0}h^{0i} - \frac{1}{2}\nabla^{2}\nabla^{2}\partial^{0}h^{0i} - \frac{1}{3}\nabla^{2}\partial^{i}\partial_{0}\partial_{0}h^{00} - \frac{1}{2}\nabla^{2}\partial^{i}\partial_{0}\partial_{k}h^{0k} \nonumber \\[5pt]
&- \frac{1}{2}\partial_{0}\partial^{0}\partial_{0}\partial^{0}\partial_{j}h^{ij} + \frac{1}{6}\partial_{0}\partial^{0}\partial_{0}\partial^{0}\partial^{i}\tilde{h} + \frac{1}{6}\nabla^{2}\partial_{0}\partial^{0}\partial^{i}\tilde{h} + \frac{1}{6}\partial_{0}\partial^{0}\partial^{i}\partial_{0}\partial_{k}h^{0k} \nonumber \\[5pt]
&+ \frac{1}{3}\partial_{0}\partial^{0}\partial^{i}\partial_{k}\partial_{l}h^{kl} - \frac{1}{2}\nabla^{2}\partial_{0}\partial^{0}\partial_{j}h^{ij}. 
\label{eq70}
\end{align}
This expression can be identified with a time derivative of $\Xi^{i}_{(2)}$ in the following way $\Xi^{i} \equiv \partial_{0}\Xi^{i}_{(2)} =  \partial_{j}\Xi^{ij}_{(8)} + \partial_{j}\Xi^{ji}_{(8)}$, which will ultimately relate $\xi_{0i}^{(2)}$ to $\xi_{(8)}^{0i}$. To introduce such time derivative we must redefine the evolution parameter $\delta\xi_{0i}^{(2)}$ as
\begin{equation}
\delta\xi_{0i}^{(2)} = \partial_{0}\delta\xi_{i}^{(2)} + \partial_{i}\delta\xi_{0}^{(2)},
\label{eq71}
\end{equation}
Inserting this into the variation of the action, and performing some partial integrations, yields 
\begin{equation}
\delta S = \int\Big[\Xi(\delta\xi^{(1)} + \delta\xi^{(6)}) - \partial_{i}\Xi^{i}_{(2)}\delta\xi_{0}^{(2)} - \Xi^{i}\left(\delta\xi_{i}^{(2)} + \delta\xi_{0i}^{(8)}\right)\Big]dtd^{3}x
\end{equation}
Where $\partial_i \Xi_{(2)}^i \delta \xi_0^{(2)}=\partial^0 \Xi \delta \xi_0^{(2)}$. With this, the expression above winds up as
\begin{equation}
\delta S = \int\left[\Xi\left(\delta\xi^{(1)} + \delta\xi^{(6)} + \partial^{0}\delta\xi_0^{(2)}\right) - \Xi^{i}\left(\delta\xi_{i}^{(2)} + \delta\xi_{0i}^{(8)}\right)\right]dtd^3x.
\label{eq72}
\end{equation}
Which in turn relates the parameters $\xi^{\mu}$ in the following way
\begin{align}
&\delta\xi^{(1)} = \delta\xi^{(6)} = \delta\xi, \label{eq73} \\[5pt]
&\partial^{0}\delta\xi_{0}^{(2)} = \partial^{0}\delta\xi_{0} = - 2\delta\xi, \label{eq74} \\[5pt]
&\delta\xi_{i}^{(2)} = \delta\xi_{i} = - \delta\xi_{0i}^{(8)}.  \label{eq75}
\end{align}
Which, when inserted back into relation (\ref{eq58}), finally leads us to the gauge symmetries of the theory
\begin{equation}
\delta h_{\mu\nu} = \eta_{\mu\nu}\delta\xi + \frac{1}{2}(\partial_{\mu}\xi_{\nu} + \partial_{\nu}\xi_{\mu})
\label{eq76}
\end{equation}
These results match those found in the literature \cite{33} where different approches were used.
}
\section{Conclusions}{
A Hamilton-Jacobi analysis for the linearized Weyl action was performed, matching previous results reported in the literature. By introducing additional variables the system was written as a first-order theory. The canonical Hamiltonian of the system was obtained, as well as the Hamiltonians, which were then classified. By using the null vectors,   this classification was performed in a consistent manner and, in this particular case, the identification of the complete group of Hamiltonians was trivial. With the help of the generalized bracket extra degrees of freedom and the non-involutive Hamiltonians were removed. This analysis was followed by the counting of the physical of degrees of freedom. The construction of the fundamental differential allowed us to identify the characteristic equations as well as a detailed identification of the gauge transformations. With this analysis performed, the theory can now be taken into the quantum realm.
}\\
Competing interests: The authors declare there are no competing interests.\\
This manuscript does not report data.


\end{document}